\documentclass{article}
\usepackage{arxiv}
\usepackage[utf8]{inputenc} 
\usepackage[T1]{fontenc}    
\usepackage{hyperref}       
\usepackage{url}            
\usepackage{booktabs}       
\usepackage{amsfonts}       
\usepackage{nicefrac}       
\usepackage{microtype}      
\usepackage{graphicx}
\usepackage{epstopdf, epsfig}
\usepackage{amsmath}
\usepackage{graphicx}
\usepackage{graphics}
\usepackage{subfig}
\usepackage{xcolor}
\usepackage{verbatim}
\usepackage{multirow}
\usepackage{enumitem}

\title{Applying deep reinforcement learning to active flow control in turbulent conditions}
\author{
  Feng Ren\\
  Research Center for Fluid Structure Interactions\\
  Department of Mechanical Engineering\\
  The Hong Kong Polytechnic University\\
   \And
  Jean Rabault\\
  Department of Mathematics\\
  University of Oslo\\
  \texttt{Contributed equally to this work} \\
   \And
  Hui Tang\\
 Research Center for Fluid Structure Interactions\\
  Department of Mechanical Engineering\\
  The Hong Kong Polytechnic University\\
  \texttt{h.tang@polyu.edu.hk}\\
}

\begin{document}
\maketitle

\begin{abstract}
Machine learning has recently become a promising technique in fluid mechanics, especially for active flow control (AFC) applications. A recent work [J. Fluid Mech. (2019), vol. 865, pp. 281-302] has demonstrated the feasibility and effectiveness of deep reinforcement learning (DRL) in performing AFC over a circular cylinder at $Re = 100$, i.e., in the laminar flow regime. As a follow-up study, we investigate the same AFC problem at an intermediate Reynolds number, i.e., $Re = 1000$, where the turbulence in the flow poses great challenges to the control. The results show that the DRL agent can still find effective control strategies, but requires much more episodes in the learning. A remarkable drag reduction of around $30\%$ is achieved, which is accompanied by elongation of the recirculation bubble and reduction of turbulent fluctuations in the cylinder wake. To our best knowledge, this study is the first successful application of DRL to AFC in weak turbulent conditions. It therefore sets a new milestone in progressing towards AFC in strong turbulent flows.
\end{abstract}

\section{Introduction}

Active flow control (AFC) is a longstanding topic in fluid mechanics. Using actuators it alters flow behavior to improve aerodynamic/hydrodynamic performance, such as lift augmentation, drag reduction, flow-induced-vibration suppression, and mixing or thermal convection enhancement. AFC can be either open-loop or closed-loop, depending on whether measured flow information is used to adjust the control. Compared with open-loop controls, well designed closed-loop controls can be adaptive and effective in a wider range of flow conditions. However, when flow is turbulent involving strong nonlinearity and multiple spatial and temporal scales, it is quite challenging to design suitable closed-loop control laws in an explicit form.

In the past few years, AFC started to benefit from advances in the field of machine learning (ML). Genetic Programming (GP) was probably the first ML technique applied in AFC. For example, \citet{gautier2015closed} applied GP to search explicit control laws for reducing the recirculation zone behind a backwards-facing step. \citet{fan2018artificial} applied the linear GP to enhance jet mixing and discovered novel wake patterns. \citet{ren2019active} adopted GP-identified control laws to successfully suppress vortex-induced vibrations in a numerical simulation environment.

Recently, a novel ML technique, deep reinforcement learning (DRL), has been attracting increasing attentions in the fluid mechanics community \citep{brunton2020machine, Rabault2020, ren2020active}, following its many successes in robotics control \citep{mnih2015human} and sophisticated game playing such as Go \citep{silver2016mastering}. Applications were mainly focused on agile maneuvering and biomimetism. For example, \citet{reddy2016learning} used DRL to train a glider to fly autonomously by exploiting thermal currents in sunny weathers. \citet{verma2018efficient} studied the locomotion of fish schoolings, and using DRL trained rear fishes to harness energy from the wake of leading fishes. In these studies, owing to the limitations of early DRL algorithms, discretized control was used, where the control space was limited to a few discrete values rather than spanning a continuous range. With continuous efforts from ML community, however, novel DRL algorithms have been developed to overcome such limitations and, in particular, the so-called ``policy-based methods" are now well suited to continuous-control problems.

By applying a policy-based method, i.e., the proximal policy optimization (PPO) method that is now regarded as one of the state-of-the-art methods for continuous control \citep{schulman2017proximal,heess2017emergence}, \citet{rabault2019artificial} achieved a drag reduction of approximately $8\%$ for a circular cylinder immersed in a laminar channel flow at a diameter-based Reynolds number $Re=100$, using a pair of anti-phase jets that are issued transversely from the top and bottom of the cylinder. To speed up the simulation-based training, \citet{rabault2019accelerating} also proposed a multi-environment approach, which opens the way to performing DRL-trained controls at higher Reynolds numbers. Following this work, \citet{tang2020robust} were able to design a robust DRL controller, which can effectively control the flow around the cylinder in the range of Reynolds numbers from 60 to 400.

In this study we aim to further push the boundary by conducting more challenging DRL-trained AFC in turbulent conditions. Specifically, we will investigate the same AFC problem as in \citet{rabault2019artificial} and \citet{tang2020robust}, but at a higher Reynolds number, i.e., $Re=1000$, where the cylinder wake becomes turbulent and difficult to control. Our results show that the DRL-trained AFC not only continues to perform well in this weak turbulent condition, but also achieves a remarkable drag reduction of around $30\%$.

\section{Methodology}
\subsection{Flow configuration}

\begin{figure}
\centerline{\includegraphics[width=9cm]{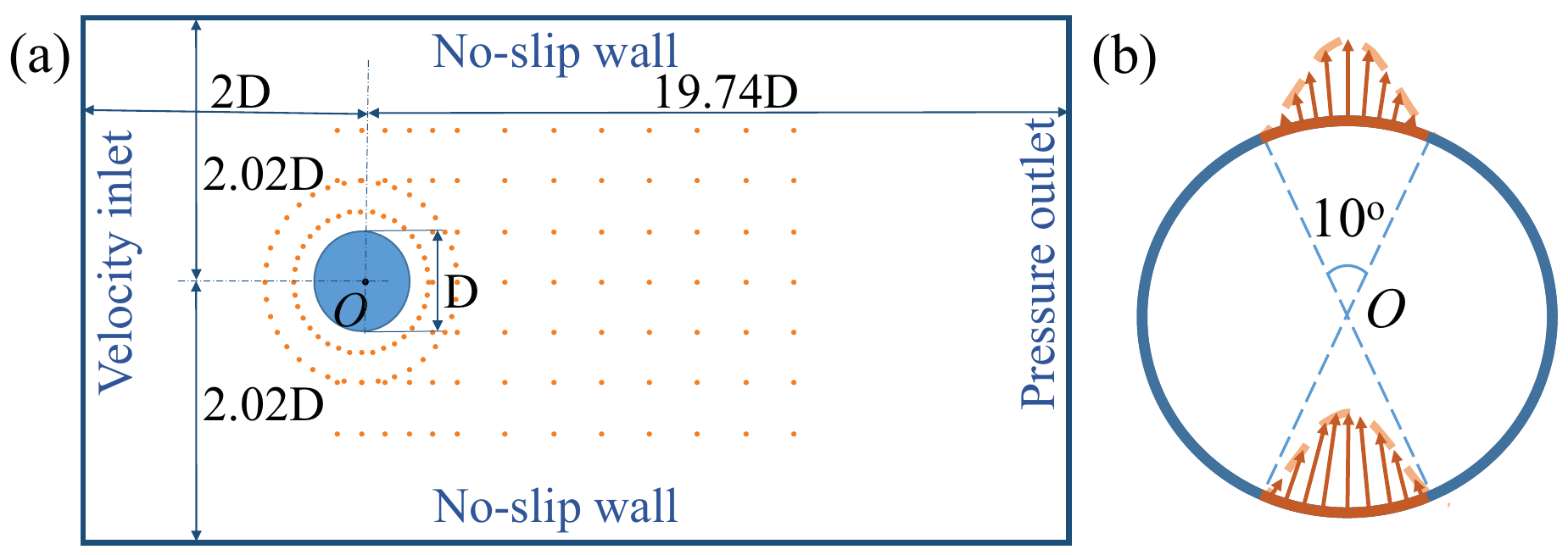}}
\caption{\label{fig:CFDsetup}Schematics of the computational domain, boundary conditions and layout of velocity sensor array (a) and the cylinder with a pair of anti-phase jets (b). A sinusoidal velocity profile is applied to the jets.}
\end{figure}

In the present work, we adopt the flow system similar to that in \citet{rabault2019artificial}, except for the Reynolds number being increased from $100$ to $1000$ to consider possible turbulence. As sketched in figure \ref{fig:CFDsetup}(a), a circular cylinder of diameter $D$ is located in the centerline of a narrow channel, $2D$ downstream of the inlet boundary and about $20D$ upstream of the outlet boundary. The incoming flow from the inlet boundary has a parabolic velocity profile. The Reynolds number is then defined based on the mean incoming velocity $U$ and the cylinder diameter $D$ as $Re = UD/\nu$, where $\nu$ is kinematic viscosity of the fluid. In the following, all results will be presented in non-dimensional form, nondimensionalized with combinations of $U$, $D$, and a reference time $T = D/U$.

An array of velocity sensors is employed to perceive the flow environment. As illustrated by orange dots around and behind the cylinder in figure \ref{fig:CFDsetup}(a), in total 151 sensors are used, each providing two time dependent signals, i.e., the streamwise and transverse velocity components.

As actuators for the flow control, a pair of jets being issued transversely are implemented at the top and bottom of the cylinder, each covering an arc of $10^{\circ}$, as depicted in figure \ref{fig:CFDsetup}(b). This jet pair operates in anti-phase, realizing a zero net mass flux at anytime. A sinusoidal profile is applied to the jet velocities, so that the no-slip boundary condition is satisfied at the slot edges. The jet centerline velocity is confined in the range of $[-1.62, 1.62]$, consistent with the range set in \citet{rabault2019artificial}.

The goal of the DRL agent is to reduce drag and meanwhile to mitigate lift fluctuation. To achieve it, we adopt a similar reward function defined in \citet{rabault2019accelerating}:
\begin{equation}\label{eq_reward}
    r=-\langle C_D \rangle_S - w {\langle |C_L| \rangle_S}
\end{equation}

\noindent where $C_D$ and $C_L$ are the drag and lift coefficients, respectively. $\langle \cdot \rangle_S$ indicates an average over a typical actuation period. $w$ is a weighting factor that weights the contributions of drag and lift fluctuation in the reward. In this study it is set as $1$, different from that used in \citet{rabault2019artificial} due to a significant increase in $\overline{|C_L|}/C_D$ from the $Re=100$ case to the $Re=1000$ case. 

\subsection{Flow solver}

In prior studies by \citet{rabault2019artificial} and \citet{tang2020robust}, the flow solver used failed at Reynolds numbers roughly larger than 500. To overcome this, here we adopt a well established lattice-Boltzmann-method (LBM) code for flow simulation. In this code, we use a uniform Cartesian mesh in the entire computational domain. The boundary conditions (BCs) are similar to those in \citet{rabault2019artificial}: a constant parabolic velocity profile is applied at the inlet and a zero pressure condition is applied at the outlet. Both BCs are implemented using the non-equilibrium extrapolation scheme \citep{zhao2002non}. The half-way bounce-back scheme \citep{he1997analytic} is used to satisfy no-penetration and no-slip BC at the top and bottom walls. As for the cylinder with jets, we apply the double linear interpolation method for curved boundary treatment \citep{yu2003viscous}, and the corrected momentum exchange method for hydrodynamic force calculation \citep{chen2013momentum}.

In ML-based AFC, it is vital to reduce the time taken by the flow solver to perform each training simulation, as many such simulations are required to find an effective control strategy. Thus, instead of conducting accurate but time-consuming direct numerical simulations (DNS) for the training, we resort to large eddy simulations (LES) when simulating flows at $Re = 1000$. In LES, fluid viscosity involves both molecular viscosity and eddy viscosity, the latter modeling subgrid-scale dissipation and being derived from a local velocity derivative tensor. In this study the Vreman model \citep{vreman2004eddy} is implemented to realize LES, where velocity derivatives are calculated using a second-order finite difference scheme. This model has been successfully implemented in our previous works \citep{ren2018lattice,ren2018gpu}.

To validate the current flow solver, we conduct simulations using different configurations and make comparisons with prior benchmark results, as summarized in table \ref{tab:meshconvergence}. For the three configurations at $Re = 100$, the intermediate mesh (Configuration II) generates $\overline{C}_D$ and $\overline{|C_L|}$ values fairly close to those obtained with the finest mesh (Configuration I). The maximum $C_D$ and $C_L$ values as well as the vortex shedding Strouhal number $St$ match the benchmark results in \citet{schafer1996benchmark} and approximate those in \citet{tang2020robust}. For the three configurations at $Re = 1000$, although LES with the coarsest mesh (Configuration VI) gives results with errors of $5\%$ in $\overline{C}_D$ and $7\%$ in $\overline{|C_L|}$ compared with the highly resolved DNS (Configuration IV), it has good numerical stability and, more importantly, takes only $2\%$ of the latter's computational time, showing a very good balance between accuracy and efficiency. Based on these results, the settings in Configurations II and VI are adopted for the DRL training at $Re = 100$ and $1000$, respectively. At $Re = 1000$, each trial simulation running for a duration of $32T$ only takes about 5 minutes, using our in-house flow solver accelerated with a NVIDIA K40c GPU. These efficient simulations to a great extent make the present DRL-based AFC feasible. On the other hand, once the training is done and effective control strategies are identified, the settings in the highly resolved DNS configuration, i.e., Configuration V, are used to evaluate the well trained PPO agents, for which the computational cost is not a big concern.

\begin{table}
\begin{center}
\caption{\label{tab:meshconvergence}Validation and convergence study}
\setlength{\tabcolsep}{1mm}{
\begin{tabular}{cccccccccc}
\hline
Re& Configuration& Method& $D/{\delta}x$& $T/{\delta}t$& $\overline{C}_D$& ${C}_{D,max}$& $\overline{|C_L|}$& ${C}_{L,max}$& $St$\\
\hline
\multirow{5}*{100}
&I   &DNS &94   &2000 &3.204 &3.244 &0.646 &1.021 &0.3021\\
&II  &DNS &47   &1000 &3.200 &3.240 &0.639 &0.999 &0.3021\\
&III &DNS &23.5 &500  &3.196 &3.236 &0.608 &0.948 &0.3030\\
&Sch{\"a}fer \emph{et al.} &DNS &&& &3.22$\sim$3.24 &&0.99$\sim$1.01 &0.295$\sim$0.305\\
&Tang \emph{et al.} &DNS &&& &3.230 &&1.032 &0.3020\\
\hline
\multirow{3}*{1000}
&IV &DNS &282  &6000 &3.476 &&2.515&&\\
&V  &DNS &141  &3000 &3.438 &&2.463&&\\
&VI &LES &70.5 &1500 &3.293 &&2.339&&\\
\hline
\multicolumn{10}{l}{\scriptsize{Note: ${\delta}x$ and ${\delta}t$ are the lattice unit length and time step, respectively, used in our LBM code.}}\\
\end{tabular}}
\end{center}
\end{table}

\begin{figure}
\centerline{\includegraphics[width=8cm]{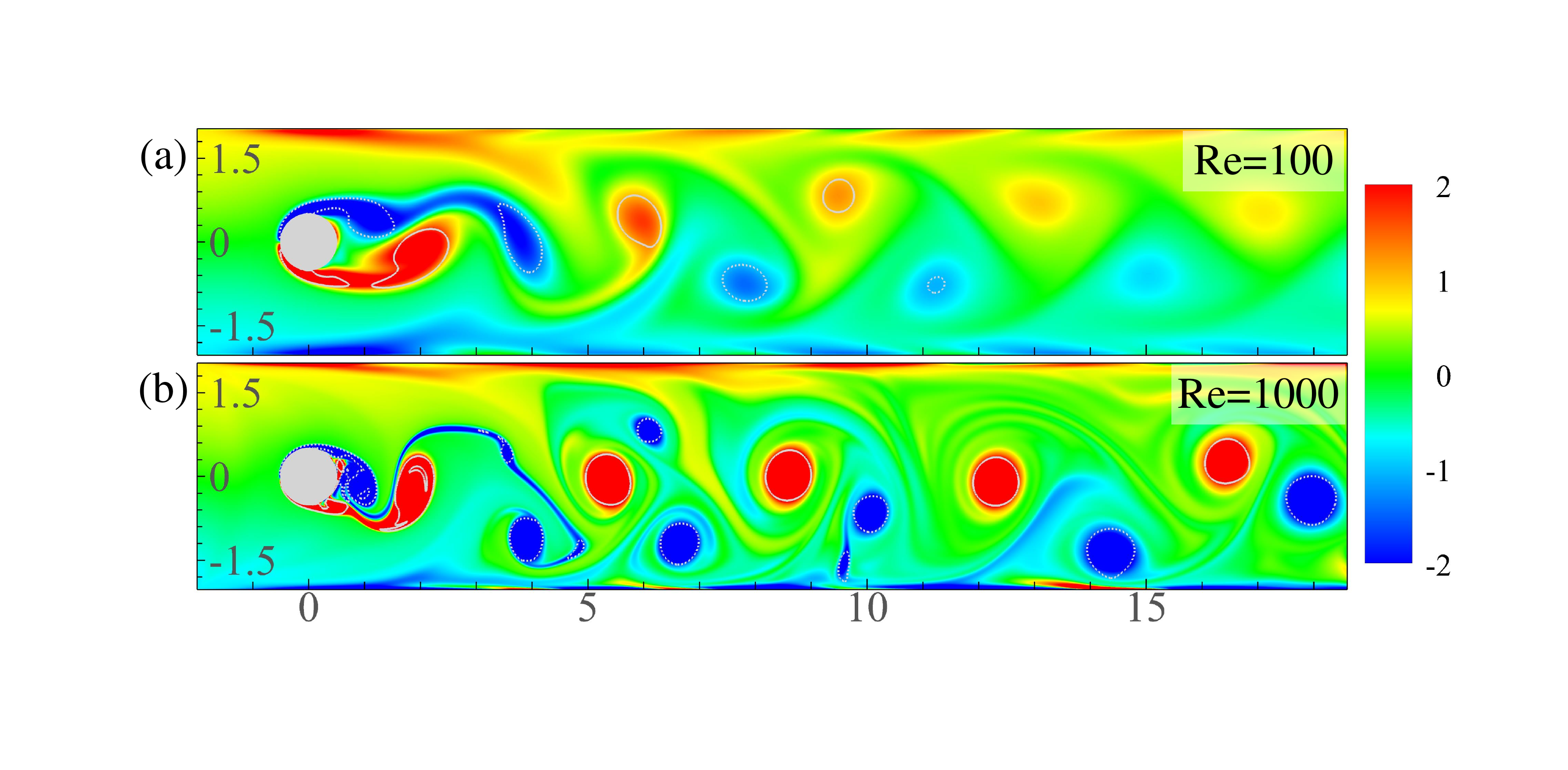}}
\caption{\label{fig:vorticity} Snapshots of the flow fields obtained at $Re = 100$ (a) and $Re = 1000$ (b). The vorticity contours are normalized by $U/D$ and scaled in a $[-2, 2]$ range. Vortices are identified using the $\lambda_{ci}$ criterion \citep{zhou1999mechanisms} and enclosed with grey lines. The chaotic features in the flow at $Re=1000$ is clearly visible.}
\end{figure}

Figure \ref{fig:vorticity} presents snapshots of typical velocity fields obtained at $Re = 100$ and $1000$. One can observe that, compared with the laminar case at $Re = 100$, the simulation at $Re = 1000$ has revealed some chaotic characteristics: vortices shed from the cylinder lose the spatial-temporal symmetry and strongly interact with the channel walls. These can also be clearly seen from the first part of the video in the supplementary material.

\subsection{Deep reinforcement learning}

\begin{figure}
\centerline{\includegraphics[width=6cm]{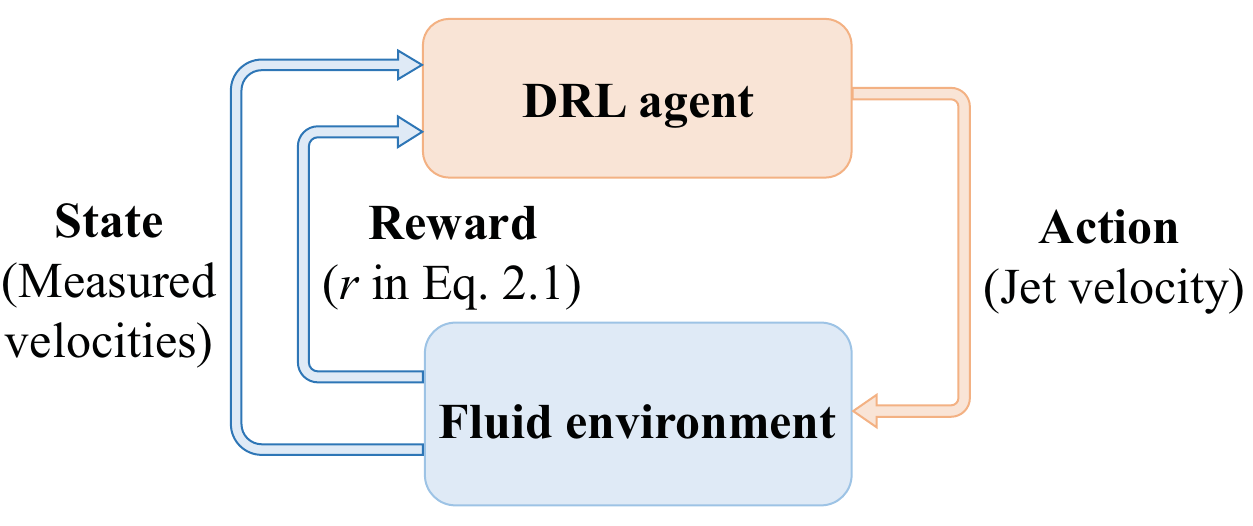}}
\caption{\label{fig:controlloop}Schematic of the DRL loop}
\end{figure}

The PPO DRL setup for performing AFC is similar to that in \citet{rabault2019artificial}. The closed-loop interaction between the cylinder, i.e., the DRL agent, and the fluid environment is depicted in figure \ref{fig:controlloop}. The sensor array collects the velocity information at selected locations, i.e., the state, from the simulation. The cylinder uses the jet-pair actuation, i.e., the action, to alter the fluid environment. The performance of the AFC is then evaluated using the reward $r$ (defined in Eq. \ref{eq_reward}). In the present study, the DRL algorithm is implemented with an in-house code using Python. More implementation details can be found in Appendix A.  

\section{Results and discussions}

To build confidence on our in-house flow solver and DRL algorithm, we start with benchmarking our DRL-trained AFC at $Re=100$ against those reported in \citet{rabault2019artificial}. Both the identified strategies and the control performance are in good agreement, as detailed in Appendix B. This constitutes an additional validation for the present methods. For the sake of brevity, in the following we only focus on discussing the DRL training at $Re=1000$. To deal with high-frequency turbulence at this $Re$, in the learning process we run each episode for $32T$ and adjust the jet actuation 5 times per $T$, both significantly longer than those adopted at $Re = 100$. 

\begin{figure}
\centerline{\includegraphics[width=12cm]{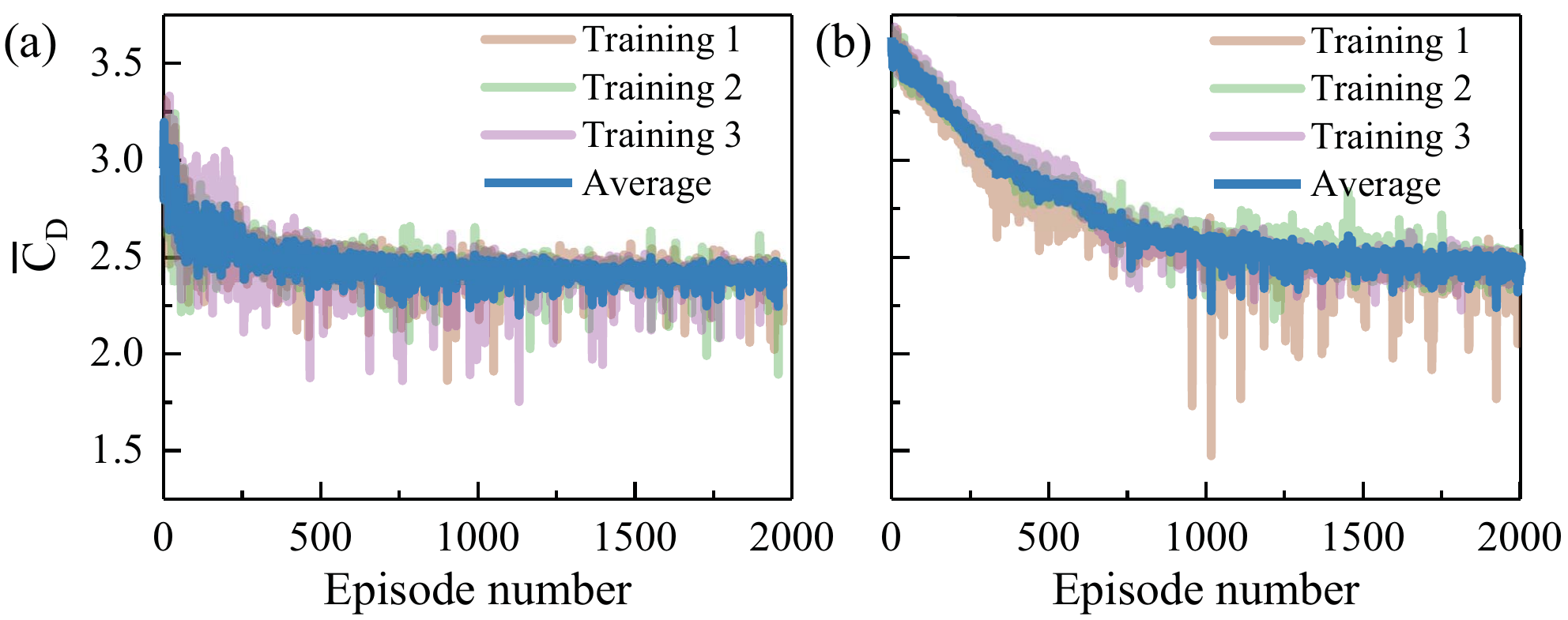}}
\caption{\label{fig:learningcurve1000}Learning curves of DRL-trained AFC at $Re = 1000$: starting from scratch (a) and starting from the strategy learnt at $Re = 100$ (b).}
\end{figure}

We adopt two different learning strategies for more challenging controls at $Re=1000$: the first one is to start the learning from a randomly initialized policy, whereas the second one is the so-called ``transfer learning", i.e., to start the learning from the well-trained policy at $Re = 100$. Learning curves are presented in figure \ref{fig:learningcurve1000}, where three independent trainings are performed for each learning strategy. Very similar learning trends are observed for the trainings under the same strategy, demonstrating the robustness of these two learning strategies. By adopting different strategies, however, an obvious difference in the learning curves is observed: the learnings using transfer learning start with much higher initial $\overline{C}_D$ values than those started from randomly initialized policies. This indicates that the policies learnt at $Re=100$ do not work well at $Re=1000$, due to the big difference in the flow dynamics. Nevertheless, all the learning curves eventually approach to similar low $\overline{C}_D$ values, mutually verifying the effectiveness of learnings using both strategies. Note that, compared to the learnings at $Re=100$, all these learnings take much more episodes to converge, revealing the difficulty in controlling chaotic flow systems.

\begin{figure}
\centerline{\includegraphics[width=7cm]{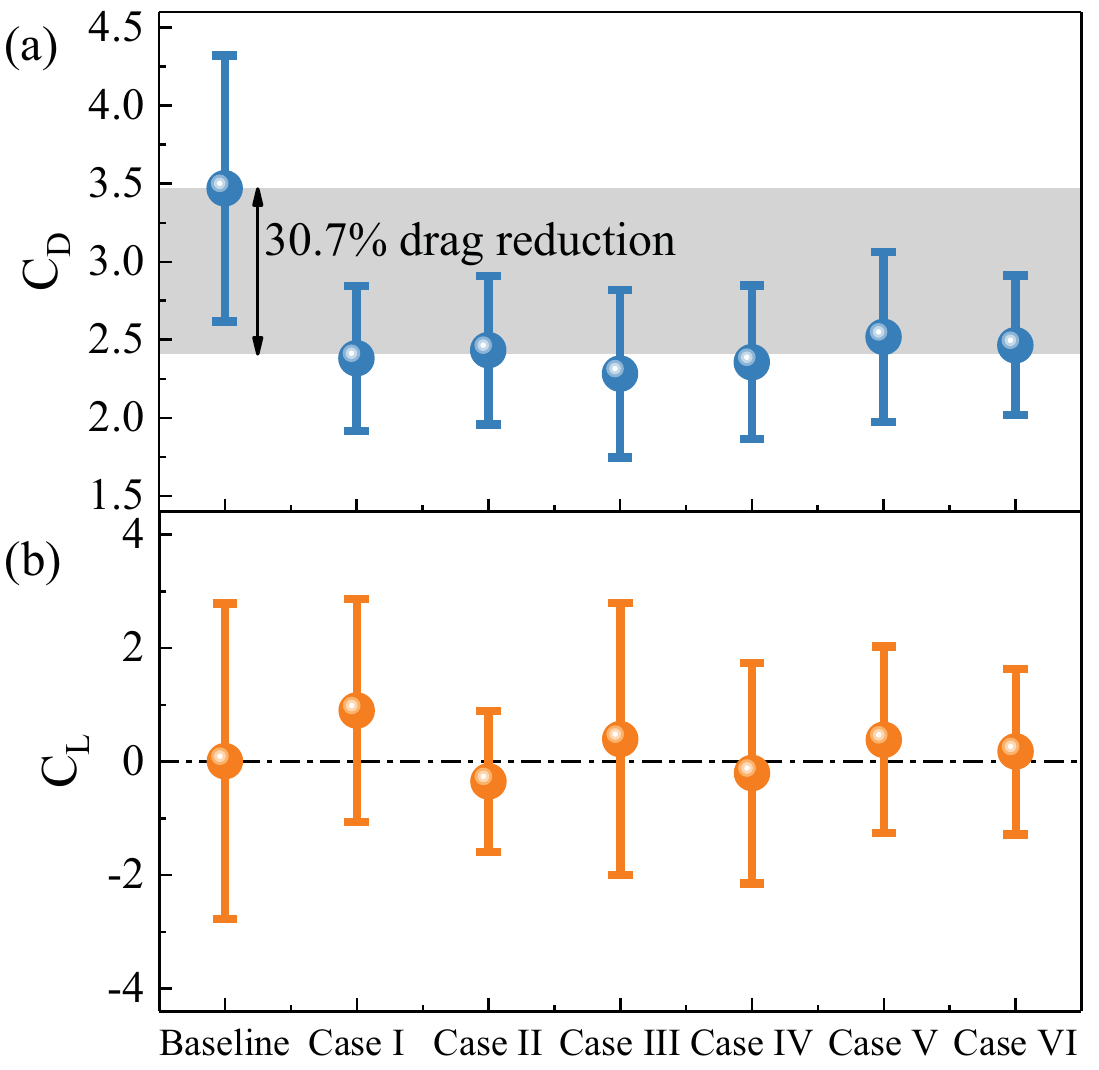}}
\caption{\label{fig:cdcl}Comparison of mean drag coefficient (a) and mean lift coefficient (b) among the baseline (i.e., uncontrolled) and six controlled cases that are evaluated using well trained policies and in deterministic mode. The error bars denote the standard deviations of the data.}
\end{figure}

Once the trainings are done, the converged policies from each training are evaluated in the deterministic mode using the highly resolved DNS configuration (i.e., Configuration V in table 1). As shown in figure \ref{fig:cdcl}(a), the evident reductions in $\overline{C}_D$ (ranging from $27.4\%$ to $34.2\%$, with a mean value $30.7\%$) predicted by DNS simulations confirm that the policies trained using less accurate LES simulations are valid. The slight variation is within expectation, which arises from the eminently random exploration mechanism present in the PPO algorithm and the strong nonlinearity and chaoticity of the turbulent flow considered here. Meanwhile, the fluctuation in $C_D$ is also greatly mitigated by the control. As revealed in figure \ref{fig:cdcl}(b), lift fluctuations are also reduced by the control. The maximum reduction of $55.2\%$ occurs in Case II. However, the control generally leads to asymmetric lift fluctuations, resulting in non-zero $\overline{C}_L$.

\begin{figure}
\centerline{\includegraphics[width=11cm]{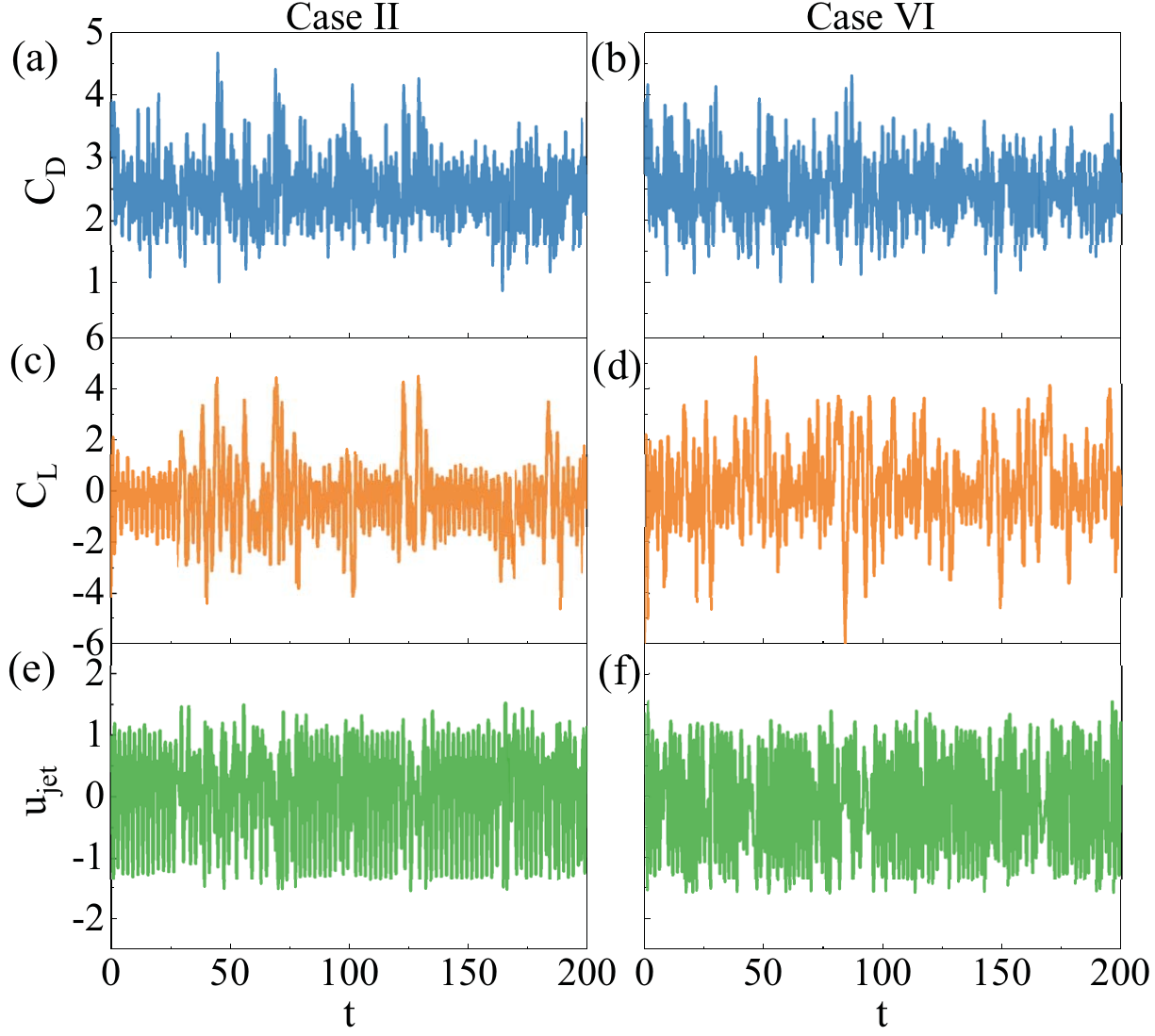}}
\caption{\label{fig:time1000}Temporal variations of drag coefficient (a-b), lift coefficient (c-d), and jet velocity (e-f) for two representative cases evaluated using well trained policies and in deterministic mode: Case DRL II (left column) and Case DRL VI (right column)}
\end{figure}

Two cases with the smallest lift fluctuations, i.e., Cases II and VI, are chosen to reveal more details about control effects of these learnt policies. The temporal variations of $C_D$, $C_L$, and jet velocity $u_{jet}$ presented in figure \ref{fig:time1000} clearly show the irregular feature of the control in turbulence conditions, contrarily to those at $Re=100$ as shown in figure \ref{fig:time100} in Appendix B. 

\begin{figure}
\centerline{\includegraphics[width=13cm]{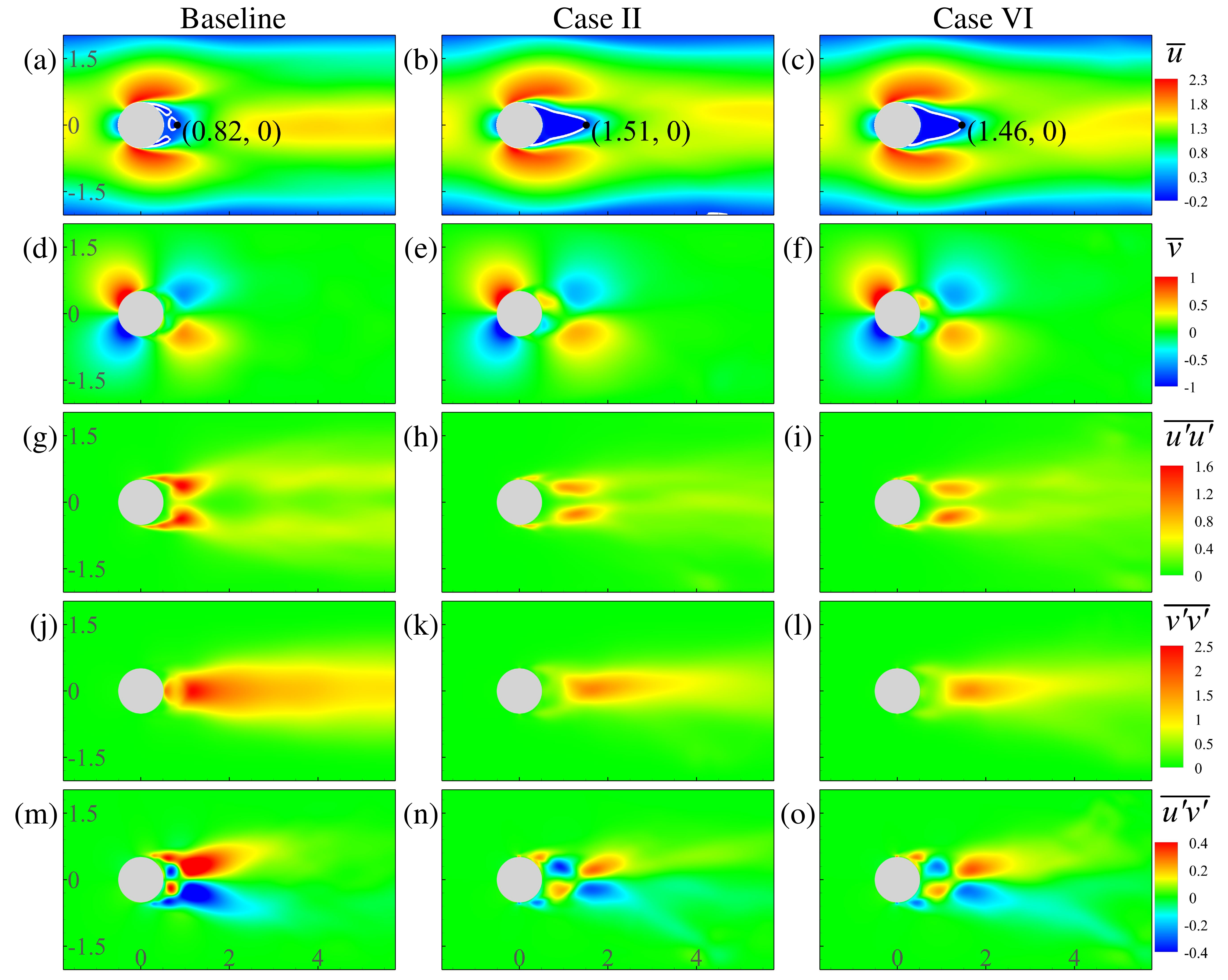}}
\caption{\label{fig:timeaveragefield} Contours of mean and turbulence quantities of the flow without (left column) and with (middle column, Case II; right column, Case VI) control: mean streamwise velocity $\overline{u}$ (a-c), mean transverse velocity $\overline{v}$ (d-f), streamwise Reynolds stress $\overline{u'u'}$ (g-i), transverse Reynolds stress $\overline{v'v'}$ (j-l), Reynolds shear stress $\overline{u'v'}$ (m-o). The white lines in (a-c) are iso-lines of zero streamwise velocity, enclosing the recirculation bubble. The data are obtained through processing $10000$ snapshots of the flow filed collected from $t = 100$ to $200$, where the control starts at $t = 0$ from a statistically stationary state.}
\end{figure}

\begin{figure}
\centerline{\includegraphics[width=12cm]{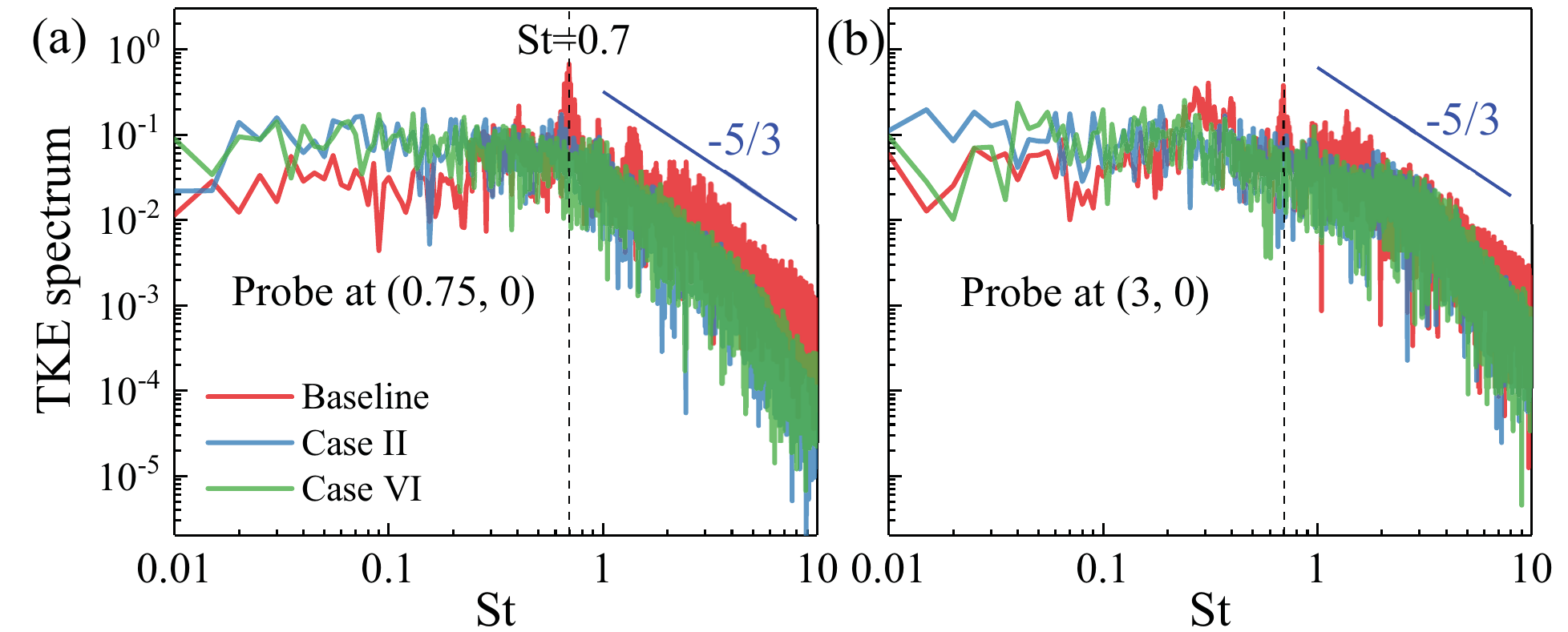}}
\caption{\label{fig:tke}TKE spectra at two locations in the centerline of the channel, i.e., $0.75D$ (a) and $3D$ (b) downstream of the cylinder center. The frequency is represented by the Strouhal number $St = fD/U$.}
\end{figure}

Contours of the first- and second-order turbulence quantities are presented in figure \ref{fig:timeaveragefield}. Comparison among the mean-streamwise-velocity contours in figure \ref{fig:timeaveragefield}(a-c) reveals that the control significantly elongates the recirculation bubble in the wake, by $211\%$ in Case II and $195\%$ in Case VI, which is accompanied by significant reduction of hydrodynamic drag as revealed earlier. The control also significantly reduces the Reynolds stresses, i.e., $\overline{u'u'}$, $\overline{v'v'}$ and $\overline{u'v'}$, in the wake region, indicating that the wake flow becomes less fluctuating. The turbulent-kinetic-energy (TKE) spectra at two selected locations in the wake, one in the recirculation bubble at $0.75D$ downstream of the cylinder center and the other outside the bubble at $3D$, are presented in figure \ref{fig:tke}. It is seen that the peak at $St\approx0.7$ in the baseline (i.e., uncontrolled) case disappears in the controlled cases, reflecting the fact that the natural vortex formation and shedding process is significantly altered by the jet actuation. In addition, with the control, obvious decrease in the spectra at higher frequencies is observed. This is consistent with the significant reduction of the Reynolds stresses revealed in figure \ref{fig:timeaveragefield}.

From the results shown in figures \ref{fig:time1000} to \ref{fig:tke}, it is seen that the flow characteristics in Cases II and VI are very similar, indicating that different learning strategies eventually give similar control policies. This similarity can also be clearly seen from the video in the supplementary material. 

\begin{figure}
\centerline{\includegraphics[width=10cm]{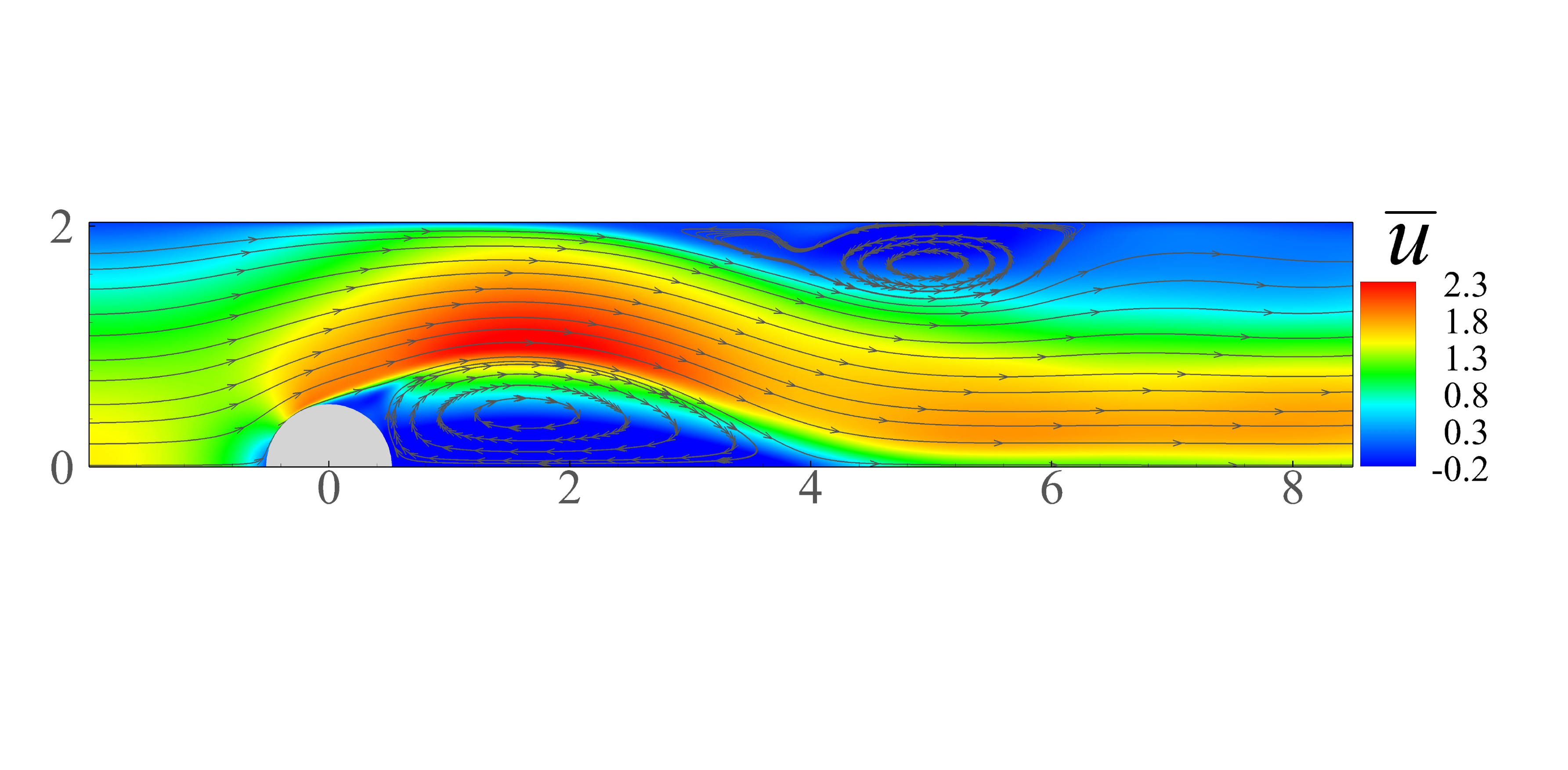}}
\caption{\label{fig:halfdomain}Time-averaged flow field of the half flow domain. The arrowed lines are streamlines, and the background is colored by mean streamwise velocity.}
\end{figure}

To further evaluate the effectiveness of the learnt control policies, we conduct a comparative study similar to that in \citet{bergmann2005optimal}, which was also adopted in, among others, \citet{rabault2019artificial} and \citet{tang2020robust}. \citet{bergmann2005optimal} suggests that the drag experienced by the cylinder consists of two main components, one arising from a ``symmetric base flow" around the cylinder, and the other arising due to the vortex shedding from the cylinder. We estimate the symmetric-base-flow drag by performing a simulation of flow around a half cylinder at $Re = 1000$ with a symmetrical boundary condition employed at the centerline of the channel. The resulting flow is shown in figure \ref{fig:halfdomain}. In this case, the mean drag coefficient on the half cylinder is read as $\overline{C}_D=0.927$. Therefore, according to \citet{bergmann2005optimal}, it is deduced that for the whole cylinder $\overline{C}_D=1.854$ if the vortex shedding is fully suppressed. This is a $47\%$ reduction from the uncontrolled $\overline{C}_D$ listed in table 1. Compare to this idealized value, the maximum reduction of $34\%$ (Case III in figure \ref{fig:cdcl}) obtained using the present DRL-trained AFC is still quite remarkable. We anticipate that further improvements can be made by using finer grained actuation, for example, through multiple jet pairs deployed on the cylinder.

\section{Summary and conclusion}

In the present work, we performed the first PPO DRL trained AFC in weak turbulent conditions, with the aim to reduce the drag and mitigate lift fluctuations experienced by a circular cylinder at $Re=1000$. The findings are summarized as follows:

\begin{itemize}
    \item At intermediate Reynolds numbers where the flow shows turbulent features, DRL can still find effective control strategies. Due to the much stronger nonlinear flow features involved, however, the learning takes much more episodes to converge than the learning in the laminar flow regime. 
    
    \item Both randomly-initialized and transfer-learning strategies perform well to reach a similar drag reduction level, i.e., around $30\%$. The latter strategy does not show obvious advantages over the former strategy as expected, mainly due to the large difference in flow dynamics caused by turbulence.
    
    \item Through analysing the AFC results, two flow features associated with the drag reduction are identified: First, the recirculation bubble is greatly elongated, similar to what has been observed in the laminar regime. Second, turbulence levels in the wake, especially in the near wake, are significantly reduced by the control.
\end{itemize}

This work further qualifies DRL as a useful machine-learning tool for solving AFC problems, and sets a new milestone by illustrating the effectiveness of DRL in a case much more complex than previous studies. We anticipate that more relevant studies will be conducted in much stronger turbulent conditions to further progress towards real-world applications.

\section{Acknowledgments}

Feng Ren and Hui Tang gratefully acknowledge financial support from the Research Grants Council of Hong Kong under the General Research Fund (Project No. 15249316 $\&$ 15214418), and the Departmental General Research Fund (Project No. G-YBXQ). They would also like to acknowledge the University Research Facility in Big Data Analytics (UBDA) of The Hong Kong Polytechnic University for providing high-performance computing resources. Jean Rabault acknowledges funding obtained through the Petromaks II project (Grand No. 280625).

\section{Appendix A: Deep reinforcement learning}

Here we present a brief description about the proximal policy optimization (PPO) algorithm. For more details, readers are referred to any of the many discussions on this topic, such as \citet{heess2017emergence}, \citet{schulman2017proximal}, and \citet{rabault2019artificial, Rabault2020}.

In each episode, the PPO agent applies the control policy $N$ times and collects a sequence of state-action-reward combinations, i.e.,
\begin{equation}
    \tau = (s_1, a_1, r_1), (s_2, a_2, r_2), (s_t, a_t, r_t),...,(s_N, a_N, r_N).
\end{equation}

To optimize against a long-term objective, the learning process is driven by a discounted reward:
\begin{equation}
    R_t = \sum_{t'>t}{\gamma^{t'-t}r_{t'}},
\end{equation}
\noindent where $0 < \gamma < 1$ is a discount factor usually close to 1, such that later rewards contribute more to the discounted reward.

\begin{figure}
\centerline{\includegraphics[width=8cm]{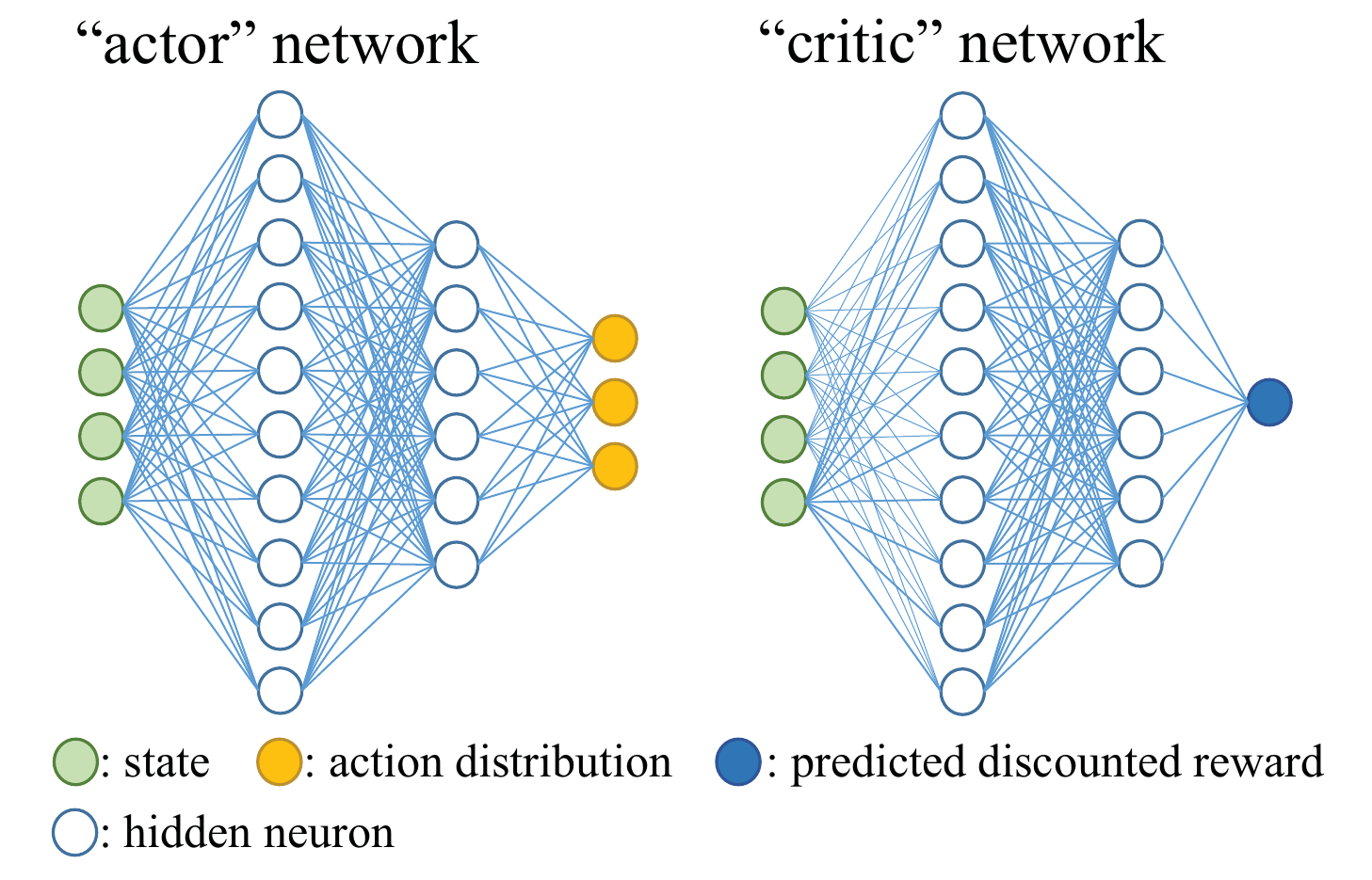}}
\caption{\label{fig:ann}General actor-critic setup used in the PPO algorithm.}
\end{figure}

The policy, $\pi_\Theta$, is modeled by an ANN having a set of weights $\Theta$. As shown in figure \ref{fig:ann}, the PPO algorithms uses two sets of ANNs: an ``actor" network whose input is the state and output the action, and a ``critic" network whose input is the state and output a prediction of the discounted reward.

In order to perform training, an appropriate loss function must be defined for each ANN. When training the critic network, an intermediate variable, i.e., the ``advantage", is used to evaluate the difference between the predicted and actual discounted rewards
\begin{equation}
    \hat{A}_t = R_t-V_\Theta(s_t).
\end{equation}
\noindent Then, the objective of the critic network is to minimize a loss function measuring the discrepancy between the predicted and actual discounted reward, i.e.,
\begin{equation}
    J_{critic} = \hat{E}_t(-\hat{A}_t^2),
\end{equation}
\noindent where $\hat{E}_t$ denotes the empirical expectation over time.

As learning progresses, the PPO agent always attempts to increase its cumulative reward. To achieve this, the actor network is used to generate actions so that the agent can interact with the environment. In return, this network is also trained using the reward information. In the present PPO implementation, we follow the work in \citet{schulman2017proximal}, where a clipped surrogate objective function is used, i.e.,
\begin{equation}
    J_{actor} = \hat{E}_t[\mbox{min}(q_t(\Theta)\hat{A}_t, \mbox{clip}(q_t(\Theta), 1-\epsilon, 1+\epsilon)\hat{A}_t)], 
\end{equation}
\noindent where $q_t(\Theta) = \pi_{\Theta}(a_t|s_t)/\pi_{old}(a_t|s_t)$ is the ratio of the probability of current policy $\pi_\Theta$ in adopting action $a_t$ according to state $s_t$ to the probability of previous policy $\pi_{old}$. The clipped term inside the above equation means that $q_t(\Theta)$ is constrained to an interval $[1-\epsilon, 1+\epsilon]$, where $\epsilon$ is a hyper-parameter set as 0.2 as recommended by \citet{schulman2017proximal}. Therefore, excessively large policy updates, which would make the training process unstable, are avoided.

When updating the policy, we use the Adam (short for ``adaptive moment estimation") optimizer, which performs better in fast convergence than conventional stochastic gradient descent optimizers \citep{kingma2014adam}. To deal with continuous control, the actor network does not directly generate actions. Instead, it generates a combination of parameters for a certain probability distribution for actions. In this study, we choose the beta distribution, from which the actions are sampled in a predefined range \citep{chou2017improving}.

Once the learning converges and the performance reaches a satisfactory level, deterministic runs can be performed to reveal more details. In these runs, the agent no longer conducts learning from the sampled data. Instead, it directly generates deterministic actions, which come with the highest probability in the distribution, i.e., no random process is involved.

\section{Appendix B: DRL control of laminar flow}

\begin{figure}
\centerline{\includegraphics[width=12cm]{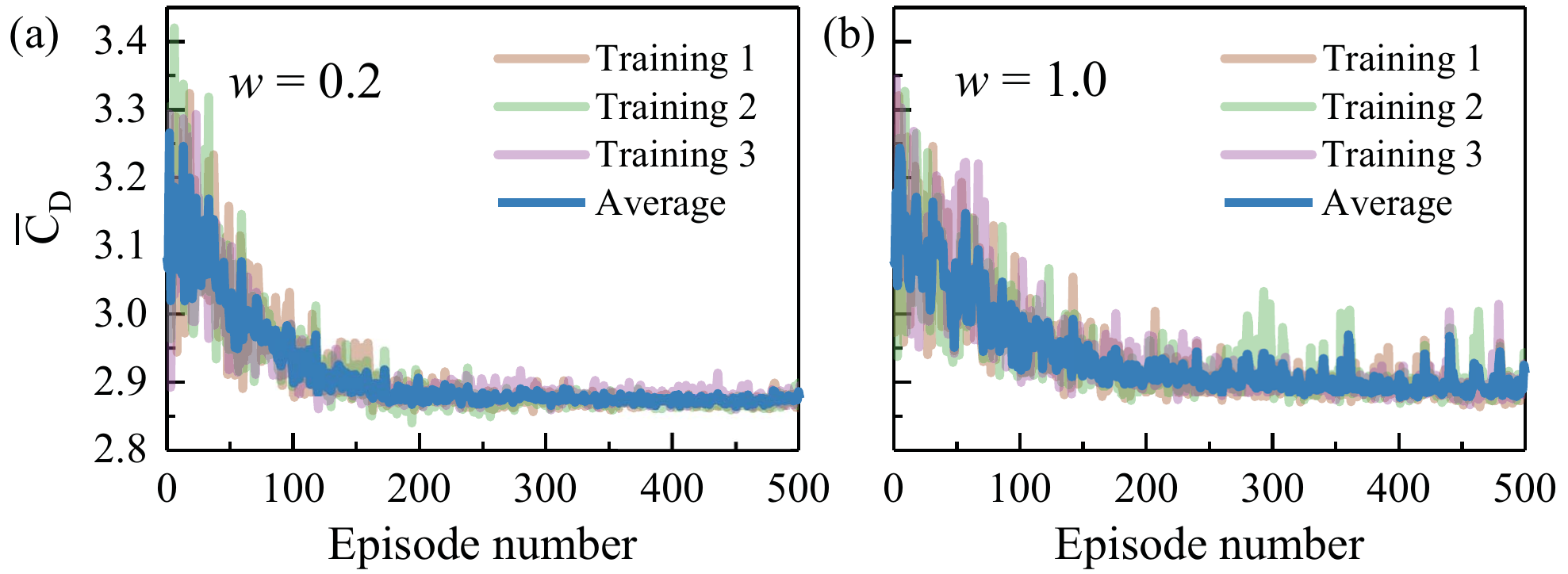}}
\caption{\label{fig:learningcurve100}Learning process at $Re = 100$: (a) using weighting factor $w = 0.2$, (b) results using $w = 1.0$.}
\end{figure}

Since in the present work different flow solver and DRL implementation are employed, we cross-validate our methods by performing the same AFC problem at $Re=100$ as in \citet{rabault2019artificial}. We adopt the same settings for the learning. The learning curves, performed using two different weighting factors for lift, i.e., $w = 0.2$ and $1.0$, are presented in figure \ref{fig:learningcurve100}, all showing good learning trends and converge within about 200 episodes. During the learning progress, each episode, i.e., a complete run of the simulation starting from a uncontrolled, fully-developed flow, runs for $24T$, corresponding to about 7.3 vortex shedding periods in the uncontrolled case. During each episode the action is adjusted 60 times according to the latest policy. To mitigate possible instability in the simulation, each adjustment of action is gradually realized using a smooth function. The control policy learnt by the PPO agent is updated every 20 episodes.

\begin{figure}
\centerline{\includegraphics[width=13cm]{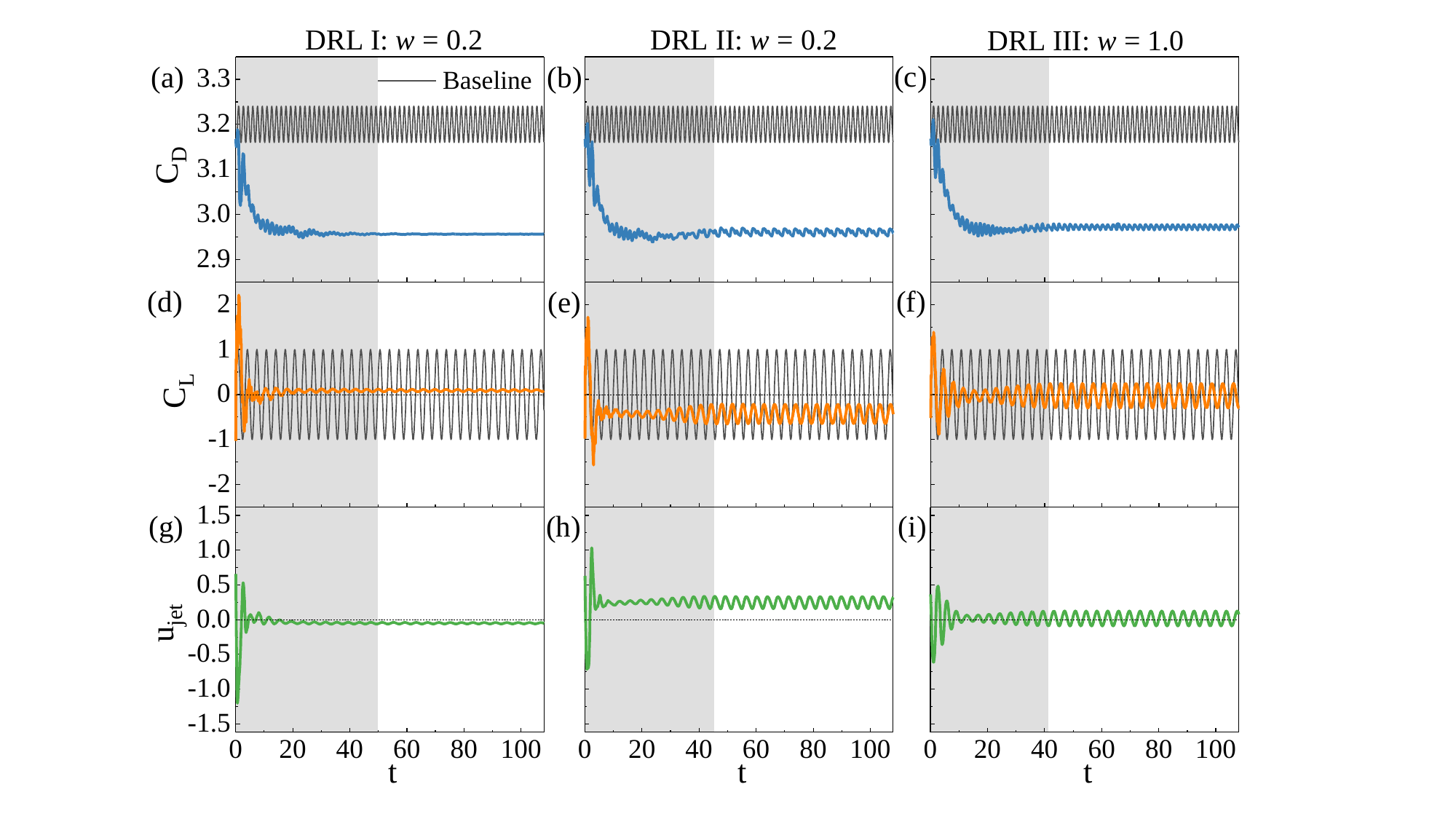}}
\caption{\label{fig:time100}Three typical control strategies observed in the deterministic run at $Re=100$. Results shown in the left and middle columns use lift weighting factor $w = 0.2$, and results shown in the right column correspond to $w = 1.0$. The black lines represent data from the baseline (uncontrolled) case. In each subfigure, the grey background indicates the transient process, whereas the white background indicates the steady-state process.}
\end{figure}

Once the training is done, deterministic controls are performed using the learnt policies. Results from three representative cases (i.e., DRL I, II and III) are presented in figure \ref{fig:time100}. The achieved drag reduction rates are $7.6\%$, $7.5\%$ and $7.1\%$, respectively, close to $8\%$ as reported in \citet{rabault2019artificial}. Note that, different weighting factors for lift give different results. As $w=0.2$ is relatively small (i.e., DRL I and II), the agent pursues larger drag reduction rates. However, as revealed in figure \ref{fig:time100}(g-h), the jet forcing is significantly biased, i.e., one side of the jet pair always blows and the other side always sucks. As a result, the cylinder will experience non-zero mean lift force, as evidenced in figure \ref{fig:time100}(d-e), which is unexpected and could cause serious problems to the structure. On the other hand, as $w=1.0$ is relatively large (i.e., DRL III), the agent weights the lift fluctuation more. Although a smaller drag reduction rate is obtained, the undesirable asymmetric lift fluctuation is significantly mitigated, as evidenced in figure \ref{fig:time100}(f). Moreover, the significant reduction in the fluctuation amplitude, compared to that in the baseline case, also suggests that the vortex formation and shedding is well suppressed by the control.

\bibliographystyle{jfm}
\bibliography{0ref}
\end{document}